\documentclass[sigconf,authorversion,nonacm]{acmart}
\usepackage{comment}
\AtBeginDocument{%
  }

\setcopyright{acmlicensed}
\copyrightyear{2025}
\acmYear{2025}
\acmDOI{XXXXXXX.XXXXXXX}
\acmConference[HUST]{Make sure to enter the correct
  conference title from your rights confirmation email}{November 16,
  2025}{St. Louis, MO USA}
\acmISBN{978-1-4503-XXXX-X/2018/06}

\begin{document}

\title{A description of the radio astronomy data processing tool DDF Pipeline}

% The "author" command and its associated commands are used to define
% the authors and their affiliations.
% Of note is the shared affiliation of the first two authors, and the
% "authornote" and "authornotemark" commands
% used to denote shared contribution to the research.
 \author{Mathis Certenais}
 \email{mathis.certenais@irisa.fr}
 \orcid{0009-0005-0659-5951}
 \affiliation{%
   \institution{Université de Rennes}
   \city{Rennes}
   \country{France}
 }

 \author{François Bodin}
 \email{francois.bodin@irisa.fr}
 \orcid{0000-0002-0928-2564}
 \affiliation{%
   \institution{Université de Rennes}
   \city{Rennes}
   \country{France}
}

 \author{Laurent Morin}
 \email{laurent.morin@irisa.fr}
 \orcid{0009-0001-4630-1234}
 \affiliation{%
   \institution{Université de Rennes}
   \city{Rennes}
   \country{France}
 }

%\renewcommand{\shortauthors}{Certenais et al.}

\begin{abstract}
This paper presents the DDF Pipeline, a radio astronomy data processing tool initially designed for the LOw-Frequency ARray (LOFAR) radio-telescope and a candidate for processing data from the Square Kilometre Array (SKA). This work describes the DDF Pipeline software and presents a coarse-grain profiling execution to characterize its performance.

\end{abstract}

\keywords{Profiling}

\maketitle
\section{Introduction}
The next-generation Square Kilometre Array (SKA) \cite{dewdney_square_2009} radio telescope is estimated to produce approximately 700 petabytes of archived data per year.
To meet the challenges posed by scientific projects that generate such large volumes of data, advanced data processing tools are essential.
The DDF Pipeline \cite{shimwell_lofar_2019, tasse_lofar_2021} is a data processing tool initially designed for the LOw-Frequency ARray (LOFAR) \cite{shimwell_lofar_2019} radio telescope and is a candidate for processing SKA data.

This work is divided into two parts: the first part describes the DDF Pipeline software, and the second part presents a coarse-grain profiling execution of the DDF Pipeline and characterizes its performance.
% \begin{itemize}
%     \item The next-generation Square Kilometre Array (SKA) \cite{dewdney_square_2009} radio telescope is estimated to produce approximately 700 petabytes of archived data per year.
%     \item Data processing tools are required to meet the challenges of scientific projects that generate large volumes of data.
%     \item The DDF Pipeline \cite{shimwell_lofar_2019, tasse_lofar_2021} is a data processing tool designed for the LOw-Frequency ARray (LOFAR) \cite{shimwell_lofar_2019} radio telescope and is a candidate for processing SKA data.
%     \item The first part of this work describes the DDF Pipeline software.
%     \item The second part presents a profiling execution of the DDF Pipeline and characterizes its performance.
% \end{itemize}

\section{Software description}
\label{section:description}
In this section, we present the DDF Pipeline \cite{shimwell_lofar_2019, tasse_lofar_2021}, a radio astronomy data processing software.
Developed by Martin Hardcastle, Tim Shimwell, Cyril Tasse and Wendy Williams \cite{hardcastle_ddf_2025}, this software was used in the production of data releases of the LOFAR Two-metre Sky Survey (LoTSS) and the Two-metre Sky Survey Deep Fields project (LoTSS Deep). The recent survey \cite{shimwell_lofar_2025} generated images from 290 terabytes of data collected from 505 hours of on-source observations spanning 7.5 years.
The DDF Pipeline is a composite application that integrates the imaging software DDFacet \cite{tasse_faceting_2018} and the calibration software killMS \cite{tasse_applying_2014, smirnov_radio_2015}. Inputs to the pipeline are datasets known as MeasurementSets (denoted MS in the remainder of the paper) \cite{schoenmakers_measurementset_2022}, and the outputs are data products, specifically sky images, in the Flexible Image Transport System (FITS) format. The DDF Pipeline operates in two main stages, each involving a sequence of calibration, imaging, and data product-generation steps. In the first stage, the pipeline processes a subset of the MS, while the second stage considers all the MS to produce the final data products. During execution with the default configuration file, the DDF Pipeline makes a total of 120 calls to killMS, processing each data subset independently, with 96 of these calls occurring during the second phase. Additionally, there are 19 calls to DDFacet: 12 calls process a subset of 6 out of the 24 data subsets, while the remaining 7 calls process the entire MS during the second stage. The default settings for the final product sizes are $20,000\times 20,000$ pixels at a resolution of 1.5 arcseconds.

\section{Profiling}
\label{section:profiling}

In this section, we present a coarse-grain profiling execution of the DDF Pipeline and characterize its performance. To assess the profiling execution of the DDF Pipeline, we utilized the Dool monitoring tool \cite{baker_scottchiefbakerdool_2024}.
The run was carried out at Eskemm Numérique \cite{the_brittany_digital_public_interest_group_eskemm_2025}, with resource usage monitored using Slurm \cite{yoo_slurm_2003}. The hardware and configuration were as follows: an HPE ProLiant DL385 Gen10 Plus (v2 model) single node equipped with 120 CPUs, 2 x AMD EPYC 7543 32-Core Processors, 32 cores each at 2.8GHz with hyper-threading, and 512 gigabytes of RAM configured with 50\% of shared memory.
We ran a Singularity container that includes the DDF Pipeline version 3.1, DDFacet version 0.7.2, and killMS version 3.1. This container version has been tested and approved by the developers. Although these application versions are several years old, their performance can be improved with the latest versions.
We executed the DDF Pipeline with the default configuration file using the L234026 dataset, which covers a frequency range between 121 MHz and 164 MHz. The MS is split into two bands: 121–129 MHz and 130–164 MHz, both with a 2 MHz resolution.
The input consisted of 24 tar-archive MeasurementSets (MS), totaling 38.72 gigabytes, which decompressed to 134.4 gigabytes. The profiling run of the DDF Pipeline was completed in 68.87 hours. The output totaled 594 gigabytes. The data products are composed of 15 images, with a total size of 17.92 gigabytes. This includes 11 full-resolution images, each of 1.575 gigabytes, and 4 low-resolution images, each of 0.148 gigabytes.

\begin{figure}
	\centering
	\includegraphics[width=1\linewidth]{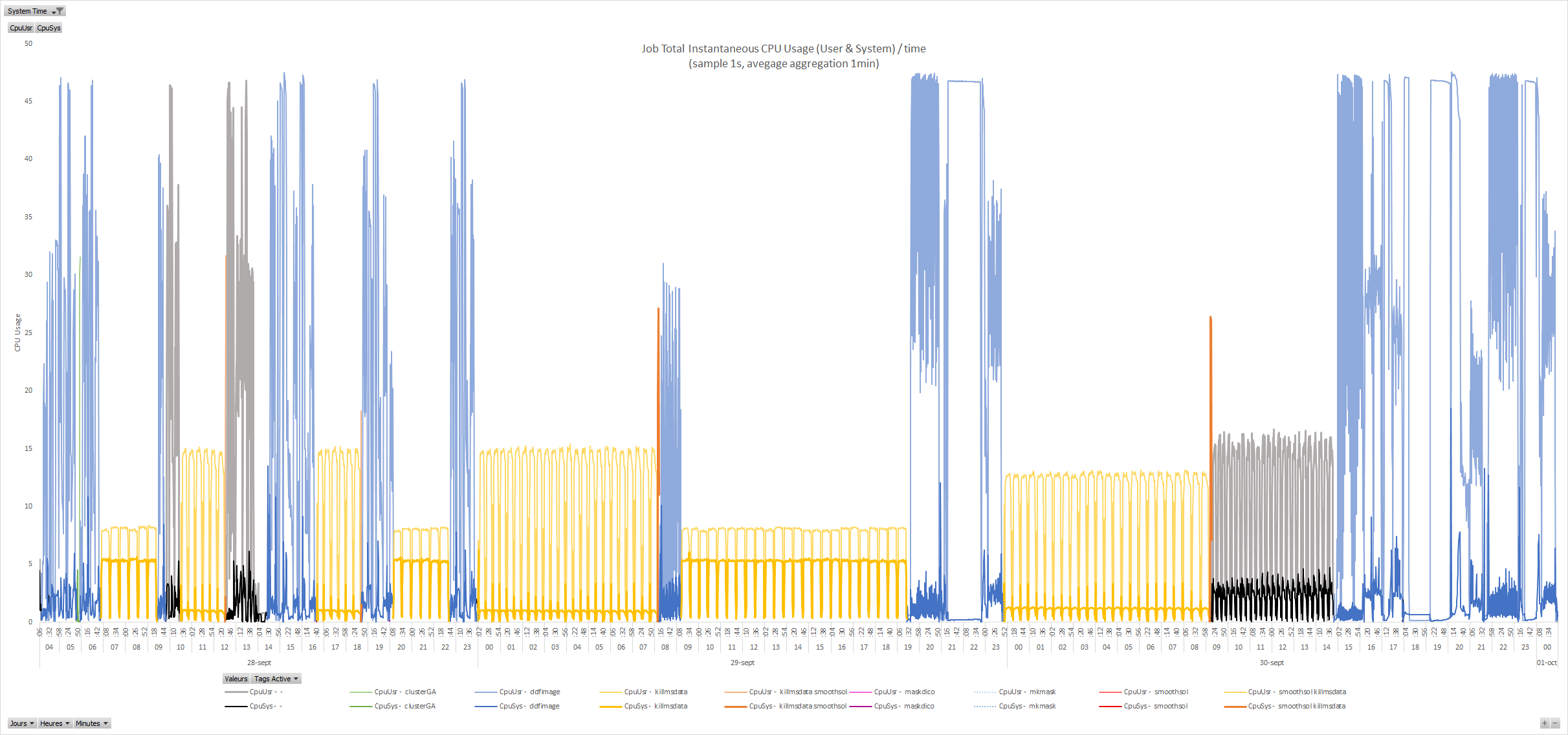}
	\caption{CPU user and system level occupancy, respectively operating system, inputs and outputs (I/Os), aggregated in average per minute, time in hours on the x-axis, number of CPUs used on the y-axis. The blue curve represents DDFacet, the yellow curve represents killMS, and the gray curve represents periods with no profile.}
	\label{fig:profiling_cpu}
\end{figure}

\begin{figure}
	\centering
	\includegraphics[width=1\linewidth]{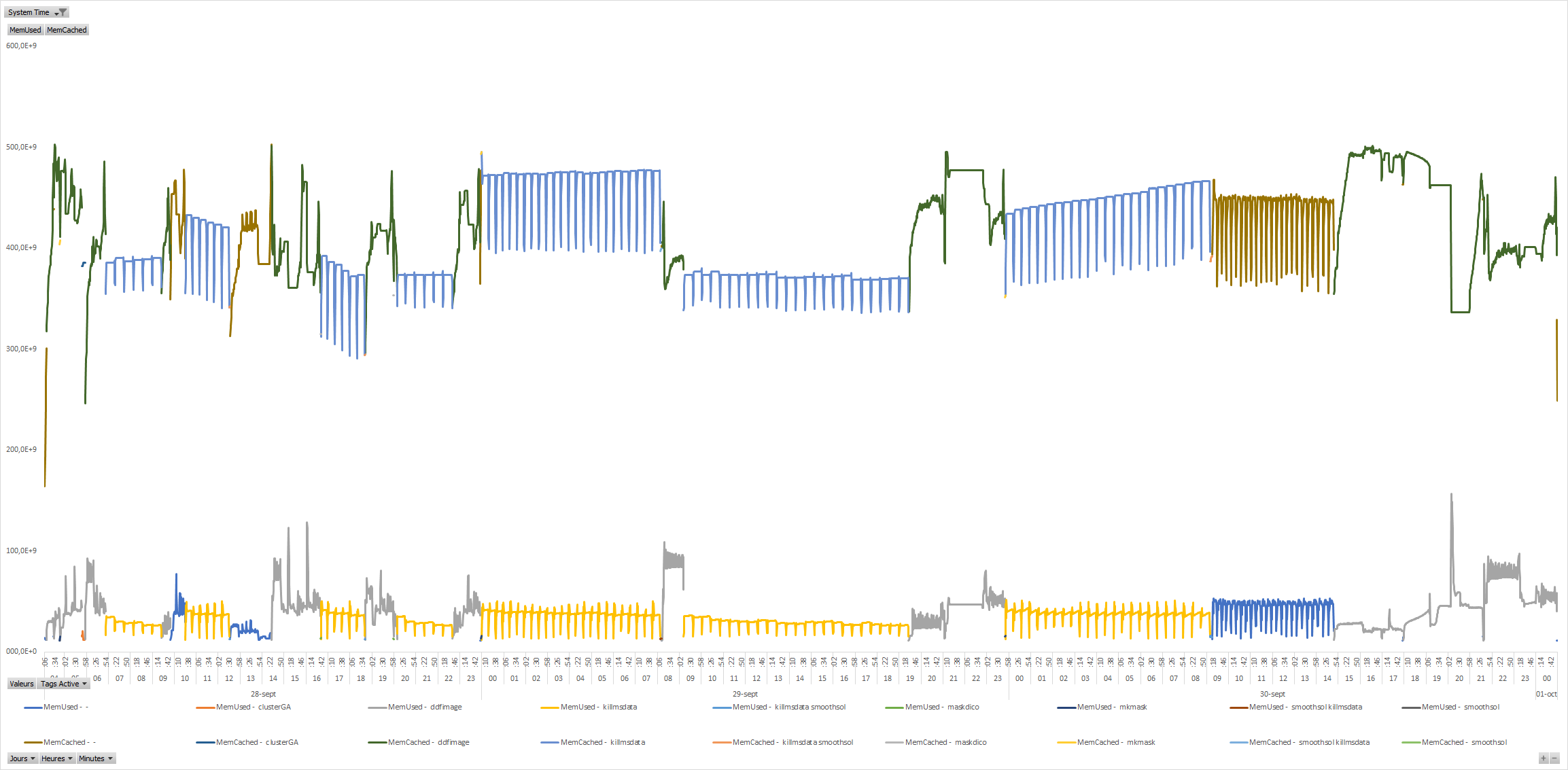}
	\caption{Memory used and cache system occupancy in average, aggregated in average per minute, time in hours on the x-axis, number of bytes on the y-axis. The blue curve represents DDFacet, the yellow curve represents killMS, and the gray curve represents periods with no profile.}
	\label{fig:profiling_ram-cache}
\end{figure}

\begin{figure}
	\centering
	\includegraphics[width=1\linewidth]{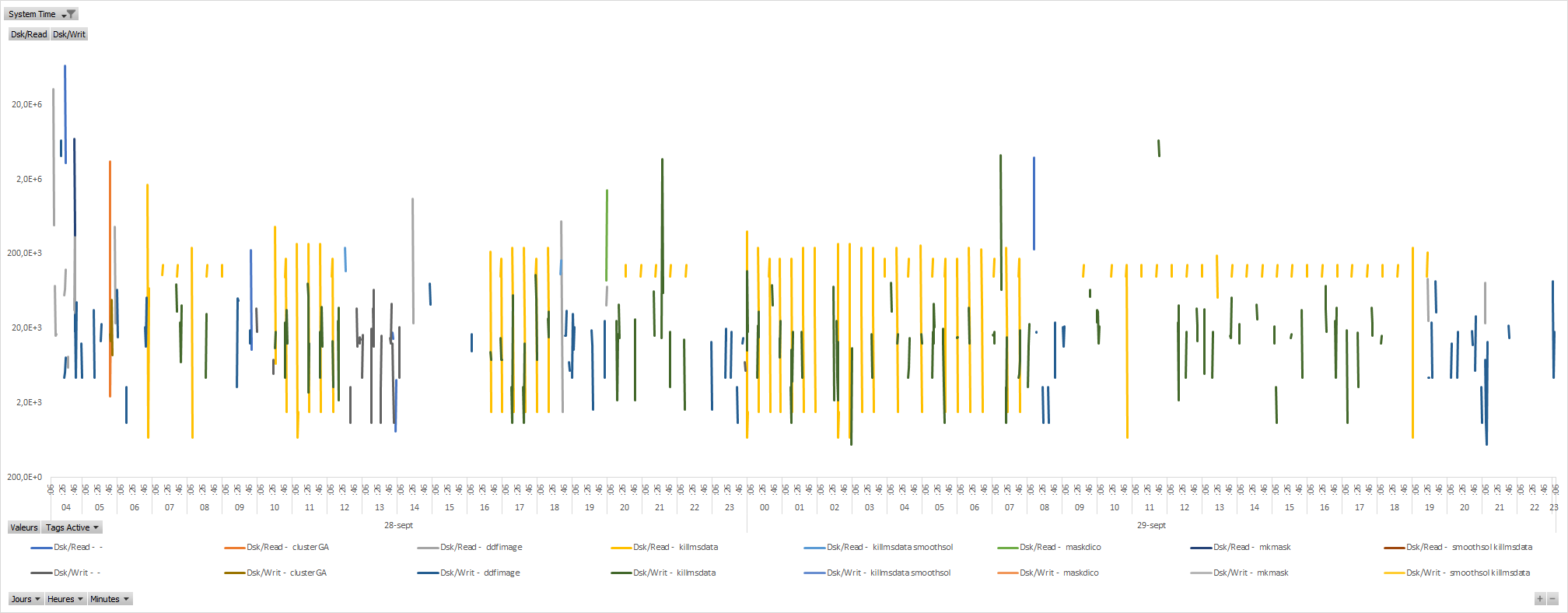}
	\caption{Disk inputs and outputs (I/Os) system occupancy, aggregated in average per minute, time in hours on the x-axis, number of I/Os in logarithmic scaling on the y-axis. The blue curve represents DDFacet, the yellow curve represents killMS, and the gray curve represents periods with no profile.}
	\label{fig:profiling_disc}
\end{figure}

\begin{table}
    \centering
    \begin{tabular}{|l|l|l|}
    \hline
        Process name & Time (s) & Time (\%) \\ \hline
        killMS & 132 388 & 53.40\% \\ \hline
        DDFacet & 83 969 & 33.87\% \\ \hline
        no profile & 29 968 & 12.09\% \\ \hline
        killMS: smoothsol 1 & 825 & 0.33\% \\ \hline
        DDFacet: clusterGA & 517 & 0.21\% \\ \hline
        DDFacet: mkmask & 157 & 0.06\% \\ \hline
        DDFacet: maskdico & 101 & 0.04\% \\ \hline
        killMS: smoothsol 2 & 4 & 0.00\% \\ \hline
        \textbf{total} & \textbf{247 929} & \textbf{100.00\%} \\ \hline
    \end{tabular}
    \caption{Duration of the main tasks executed during the DDF Pipeline, including the time taken in seconds and the percentage of total execution time for each process.}
    \label{fig:table-time}
\end{table}

\begin{table}
    \centering
    \begin{tabular}{|l|c|c|c|c|}
        \hline
        \multicolumn{1}{|c|}{Process name} & \multicolumn{2}{c|}{CPU} & \multicolumn{2}{c|}{Memory} \\ 
         & Usr & Sys & Used (GB) & Cached (GB) \\ 
        \hline
        DDFacet & 25.1 & 1.5  & 45.7 & 428.5 \\ 
        \hline
        no profile & 12.3 & 1.9 & 37.4 & 425.2 \\ 
        \hline
        killMS: smoothsol 1 & 10.9 & 12.0 & 13.6 & 382.6 \\ 
        \hline
        killMS & 9.2 & 2.5 & 33.6 & 414.4 \\ 
        \hline
        DDFacet: clusterGA & 5.5 & 0.7 & 14.1 & 384.4 \\ 
        \hline
        DDFacet: maskdico & 0.7 & 0.1 & 13.0 & 376.5 \\ 
        \hline
        killMS: smoothsol 2 & 0.7 & 0.1 & 11.7 & 354.4 \\ 
        \hline
        DDFacet: mkmask & 0.5 & 0.7 & 14.8 & 388.3 \\ 
        \hline
    \end{tabular}
    \caption{Resource occupancy information of the CPU usage (User and System) and memory usage in gigabytes during the DDF Pipeline run}
    \label{fig:table-occupancy}
\end{table}

\begin{table}
    \centering
    \begin{tabular}{|l|c|c|}
        \hline
        \multicolumn{1}{|c|}{Process name} & \multicolumn{2}{c|}{Disk}  \\ 
        & \#Total Read & \#Total Write   \\ 
        \hline
        DDFacet & 5488.3 & 13034.0  \\ 
        \hline
        no profile &  68923.8 & 6335.8  \\ 
        \hline
        killMS: smoothsol 1 &  63353.5 & 618.8 \\ 
        \hline
        killMS &  11641.1 & 18397.8   \\ 
        \hline
        DDFacet: clusterGA & 125542.0 & 10711.4   \\ 
        \hline
        DDFacet: maskdico &  294100.9 & 15248.5  \\ 
        \hline
        killMS: smoothsol 2 & 1241088.0 & 0.0  \\ 
        \hline
        DDFacet: mkmask & 647272.4 & 55726.5  \\ 
        \hline
    \end{tabular}
    \caption{Resource occupancy information of the disk read and write operations during the DDF Pipeline run}
    \label{fig:table-occupancy-disk}
\end{table}

\section{Conclusions and Future works}
In this work, we described and characterized the performance of the DDF Pipeline through a coarse-grain profiling execution. Our execution of the DDF Pipeline, utilizing the Dool monitoring tool and running on an HPE ProLiant DL385 Gen10 Plus single node at Eskemm Numérique, was completed in 68.87 hours. The pipeline processed 24 tar-archive MeasurementSets (MS) totaling 134.4 gigabytes and produced 594 gigabytes of output.
Future work will focus on porting the DDF Pipeline to the Jean Zay HPC center.

% \begin{itemize}
%     %\item This work allowed us to study and execute the DDF pipeline, and demonstrate its use as a case study for Cross-Facility Workflows deployment, which refer to scientific applications that span multiple infrastructures including instruments, data centers and HPC centers.
%     \item % Donner une premiere estimation des perfomances caracterisant l'execution de DDF Pipeline
%     \item Portage of the DDF Pipeline on Jean Zay HPC center, and work on distribution of calculations with MPI. Understand and reduce memory footprint.
%     \item Future work will focus on DDF Pipeline's deployment and execution in IDRIS HPC center. Deploy latest versions (distributed) of Singularity image softwares DDF Pipeline, killMs and DDFacet.
%     \item % Results DDN I/O, ask if I can integrate one graph.
%     \item \textit{This work has been funded by the Exa-AToW \cite{pepr_numpex_exa-atow_2025} /NumPEx Program.}
% \end{itemize}

\bibliographystyle{ACM-Reference-Format}
\bibliography{references/main}

\end{document}